\documentclass[conference]{IEEEtran}

\ifCLASSINFOpdf
\usepackage[pdftex]{graphicx}
\else
\fi

\usepackage{balance}
\usepackage{cite}
\usepackage{graphicx}
\usepackage{float}
\usepackage{amsmath}
\usepackage[utf8]{inputenc}
\usepackage[final]{pdfpages}
\usepackage{pdfpages}
\usepackage{lipsum}
\usepackage{textcase}
\usepackage{url}
\usepackage{amsmath,esint}
\usepackage{epstopdf}
\usepackage{array}

	\newcommand\blfootnote[1]{%
		\begingroup
		\renewcommand\thefootnote{}\footnote{#1}%
		\addtocounter{footnote}{-1}%
		\endgroup
	}
\newcolumntype{P}[1]{>{\centering\arraybackslash}p{#1}}
\newcolumntype{M}[1]{>{\centering\arraybackslash}m{#1}}

\pdfoutput=1

\begin{document}

\title{UAV Air-to-Ground Channel Characterization for mmWave Systems}

\author{
\IEEEauthorblockN{Wahab Khawaja, Ozgur Ozdemir, and 
Ismail Guvenc}
\IEEEauthorblockA{Department of Electrical and Computer Engineering, North Carolina State University, Raleigh, NC}
Email: \{wkhawaj, oozdemi, iguvenc\}@ncsu.edu
	
}

\maketitle
\blfootnote{This work has been supported in part by the National Science Foundation under the grants CNS-1618692 and CNS-1453678, and W. Khawaja has been supported via a Fulbright scholarship.}

\begin{abstract}
Communication at mmWave bands carries critical importance for 5G wireless networks. In this paper, we study the characterization of mmWave air-to-ground (AG) channels for unmanned aerial vehicle (UAV) communications. In particular, we use ray tracing simulations using Remcom Wireless InSite software to study the behavior of AG mmWave bands at two different frequencies: 28~GHz and 60~GHz. Received signal strength (RSS) and root mean square delay spread (RMS-DS) of multipath components (MPCs) are analyzed for different UAV heights considering four different environments:  urban, suburban, rural, and over sea. It is observed that the RSS mostly follows the two ray propagation model along the UAV flight path for higher altitudes. This two ray propagation model is affected by the presence of high rise scatterers in urban scenario. Moreover, we present details of a universal serial radio peripheral (USRP) based channel sounder that can be used for AG channel measurements for mmWave (60 GHz) UAV communications.  


\end{abstract}

\IEEEpeerreviewmaketitle

\section{Introduction}
Unmanned aerial vehicles (UAVs) are 
envisioned to support numerous applications in 5G wireless networks~\cite{3GPP_UAV,Qualcomm_UAV}. They can serve as mobile \emph{hot spots} in congested areas~\cite{rupasinghe2016optimum}, help in improving the  quality of experience of wireless users by serving as a content cache~\cite{chen2017caching}, and support communication needs of first responders at times/locations where most needed~\cite{merwaday2016improved}.  
Due to their agile, three dimensional mobility in free space, UAVs can easily move from one place to another in order to provide on demand communication support, can change their locations to avoid blockage/shadowing, and can help in data relaying/ferrying applications. 

Use of mmWave bands can be very promising for UAV communications to sustain data rate demands for high throughput mobile applications.  In particular, UAVs can maintain line-of-sight (LOS) connectivity (or at least  a reasonable non-LOS (NLOS) link) with a desired user by hovering at a favorable location~\cite{rupasinghe2016optimum}, which is crucial in maintaining a good link quality at mmWave bands due to large path loss. A number of licensed/unlicensed mmWave bands are potentially available for use in UAV communications: see e.g. FCC's \emph{spectrum frontiers} report and order for bands above 24~GHz~\cite{FCC2}, and  FCC's amendment for use of 57-60~GHz unlicensed band for outdoor communications~\cite{FCC}. In this paper, we consider the 28~GHz licensed bands and the 60~GHz unlicensed bands for their potential use in mmWave UAV communications. While propagation characteristics for both mmWave bands have been studied for urban scenarios in the literature~\cite{akdeniz2014millimeter,Lit1,Lit2,Lit3}, and air-to-ground (AG) propagation measurements are recently reported for the UWB spectrum below 6~GHz~\cite{khawaja2016uwb}, to our best knowledge characteristics of the AG UAV links have not been studied yet for the mmWave bands. 


In this paper, we analyze AG propagation characteristics for the 28~GHz and 60~GHz mmWave bands using ray tracing simulations. The analysis is carried out considering four different environments: urban, suburban, rural, and over sea. Each scenario has different number of scatterers at different heights. The transmitter (ground station (GS)) location is at a fixed position, while the UAV that carries the receiver node flies on a linear path.  A single, omni-directional antenna is used for both the transmitter and the receiver. The simulations are run for different UAV heights for each scenario, considering both 28~GHz and 60~GHz bands. The velocity of the UAV is fixed at 15~m/s. Due to high path loss at mmWave bands,  flight distance was limited to 2~km. 

Based on the simulation results it is observed that the received signal strength (RSS) reasonably follows the two ray propagation model~\cite{matolak2017air,sun2017air}. On the other hand, the two ray propagation model holds better at shorter transmitter-receiver separations; after some distance, only the dominant component is the main contributor to the RSS, and reflected path from the ground is negligible. The reflections and diffractions from the scatterers in the environment introduce fluctuations in the RSS that are dependent on the density and height of the scatterers. In case of large number of high-rise scatterers such as in the urban environment, there are rapid fluctuations in the RSS, and the two ray propagation model cannot be directly applied. The effect of scatterers is higher at 28~GHz compared to 60~GHz as can be observed from the root mean square delay spread (RMS-DS) of multipath components (MPCs). 

In addition to ray tracing simulations, we have built an AG mmWave channel sounder using universal serial radio peripheral (USRP) X310 software defined radios (SDRs), and frequency up-converters from Pasternack. The in-phase (I) and quadrature (Q) components from the USRP X310 are up-converted to 57-64~GHz unlicensed band. A basic channel sounder with frequency offset correction is built using GNU radio, and associated block diagrams for channel sounding are provided. Due to compact size of the setup, transmitter/receiver can be easily air lifted with moderate payload capacity UAVs. Our future work includes AG measurements with DJI S-1000 UAV using this sounder setup. 

The rest of this paper is organized as follows. In Section~\ref{Sec:Sec2}, ray tracing simulations for AG channel characterization in mmWave bands is described. Section~\ref{Sec:Sec3} introduces the details of the SDR based channel sounder at the 60~GHz band. Finally, the last section provides some concluding remarks.  

\section{Ray Tracing Simulations for mmWave UAVs}\label{Sec:Sec2}

Ray tracing can provide a deterministic way of characterizing the mmWave channel in different scenarios. Due to challenges involved in AG channel measurements at mmWave bands using UAVs, ray tracing offers a convenient alternative way of evaluating the channel behavior. In this section, we use the Remcom Wireless InSite ray tracing software, and imitate the real time motion of the UAV over a given trajectory. We will first describe the different ray tracing scenarios considered, followed by RSS and RMS-DS results. 



\begin{figure}[t]
	\centering
	\includegraphics[width=\columnwidth]{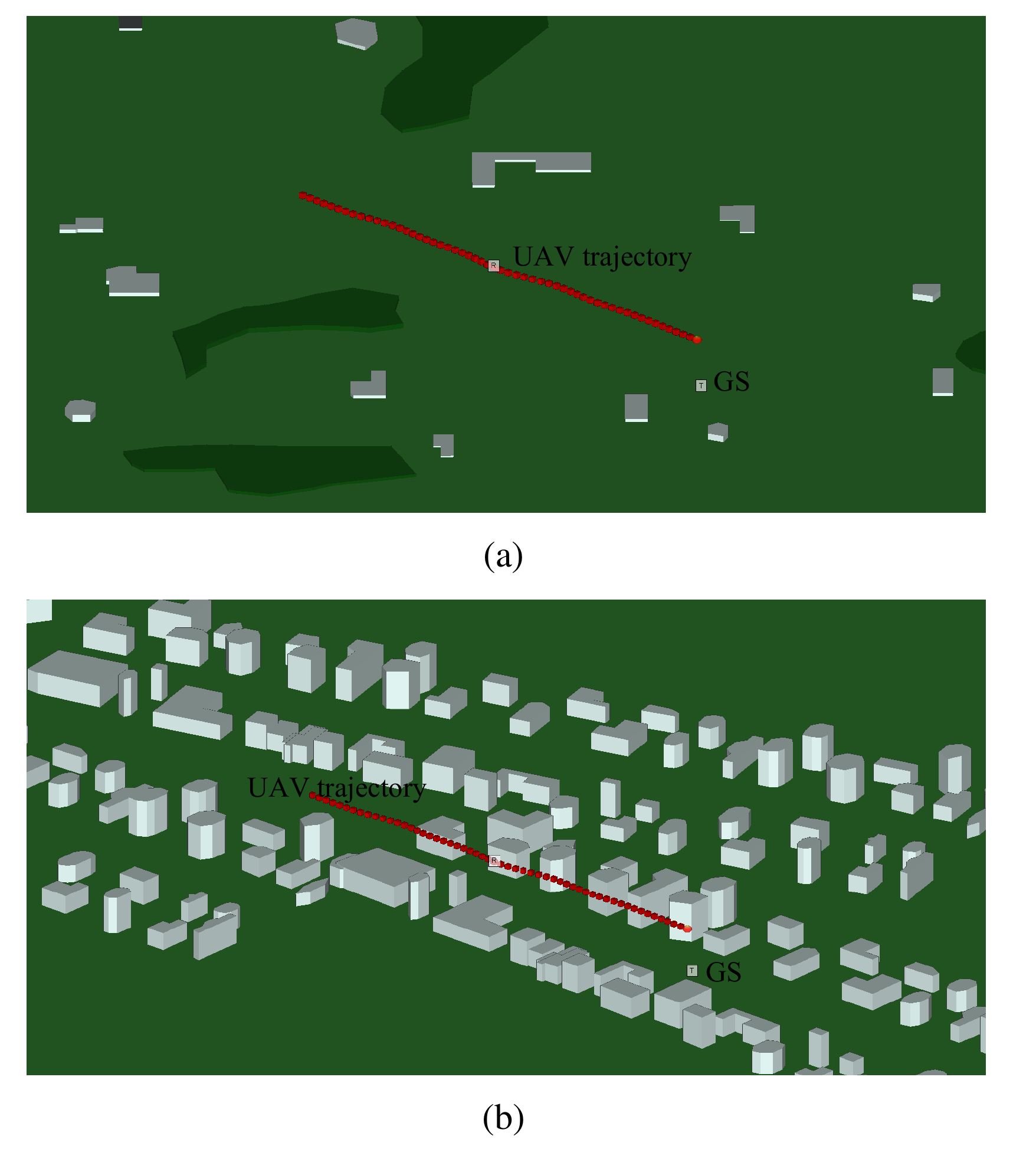}
	    \caption{UAV AG propagation scenarios simulated using Remcom Wireless InSite software. Transmitter is located on the ground, and UAV with the on-board receiver follows the illustrated trajectory. (a) Sub-urban scenario, (b) Urban scenario.}\label{Fig:setup}
\end{figure}

\begin{table}[!h]
	\begin{center}		
		\caption{Ray tracing simulation parameters.}\label{Table:Table_RT}
        	\begin{tabular}{|P{2cm}|P{2.5cm}|P{2.8cm}|}
			\hline
			\textbf{Scenario}&\textbf{Building height (m)}&\textbf{Number of buildings}\\
			\hline
			Over sea &-& - \\
            \hline
            Rural & 4-8&10\\
            \hline
            Suburban & 4-30&20 \\
            \hline
            Urban & 70-180&100 \\
            \hline
            
		  \end{tabular}
		\end{center}
 \end{table}

\subsection{Ray Tracing Scenarios}

We consider four different scenarios for ray tracing simulations: urban, suburban, rural, and over sea.
Three key factors for AG ray tracing simulations in a given scenario are the number, material, and height of the scatterers in the environment. The largest scatterers in AG propagation are the buildings. The effect of mobile vehicles on road as potential scatterers is negligible in AG propagation. We considered different number of buildings at different heights in the simulations as summarized in Table~\ref{Table:Table_RT}. The urban scenario has the highest density and height of randomly shaped buildings followed by suburban and rural. Additionally, there is scattered foliage introduced in the suburban and rural scenarios. The material of the buildings is frequency sensitive concrete at 28~GHz and 60~GHz for all the scenarios except the simulations over the sea. For simplicity, we considered the building material of all the buildings as concrete. For all the land based simulations, a dry ground is considered. Fig.~\ref{Fig:setup} shows the suburban and urban scenarios with the GS on the ground, and UAV trajectory at is shown at a height of 150~meters above the ground. In case of over sea scenario, a 10 meter layer of sea water is considered covering a typical terrain. A couple of metallic ships on the surface of the sea water are introduced as scatterers. 

The characteristics of the simulation set up are as follows. The dimensions of the terrain are 10~km by 10~km. The height of the transmitter is 2~m and is fixed on ground. The UAV heights are 2~m (imitating a mobile vehicle on the ground), 50~m, 100~m, 150~m above the ground, with a velocity of 15~m/s for all cases. The length of the UAV trajectory is around 2~km. Half-wave dipole antennas with vertical polarization are used on both transmitter and the receiver. A sinusoid is used to sound the channel at two center frequencies of $28$~GHz and $60$~GHz. The transmit power is set at 30~dBm.

\begin{figure}[t]
	\centering
	\includegraphics[width=\columnwidth]{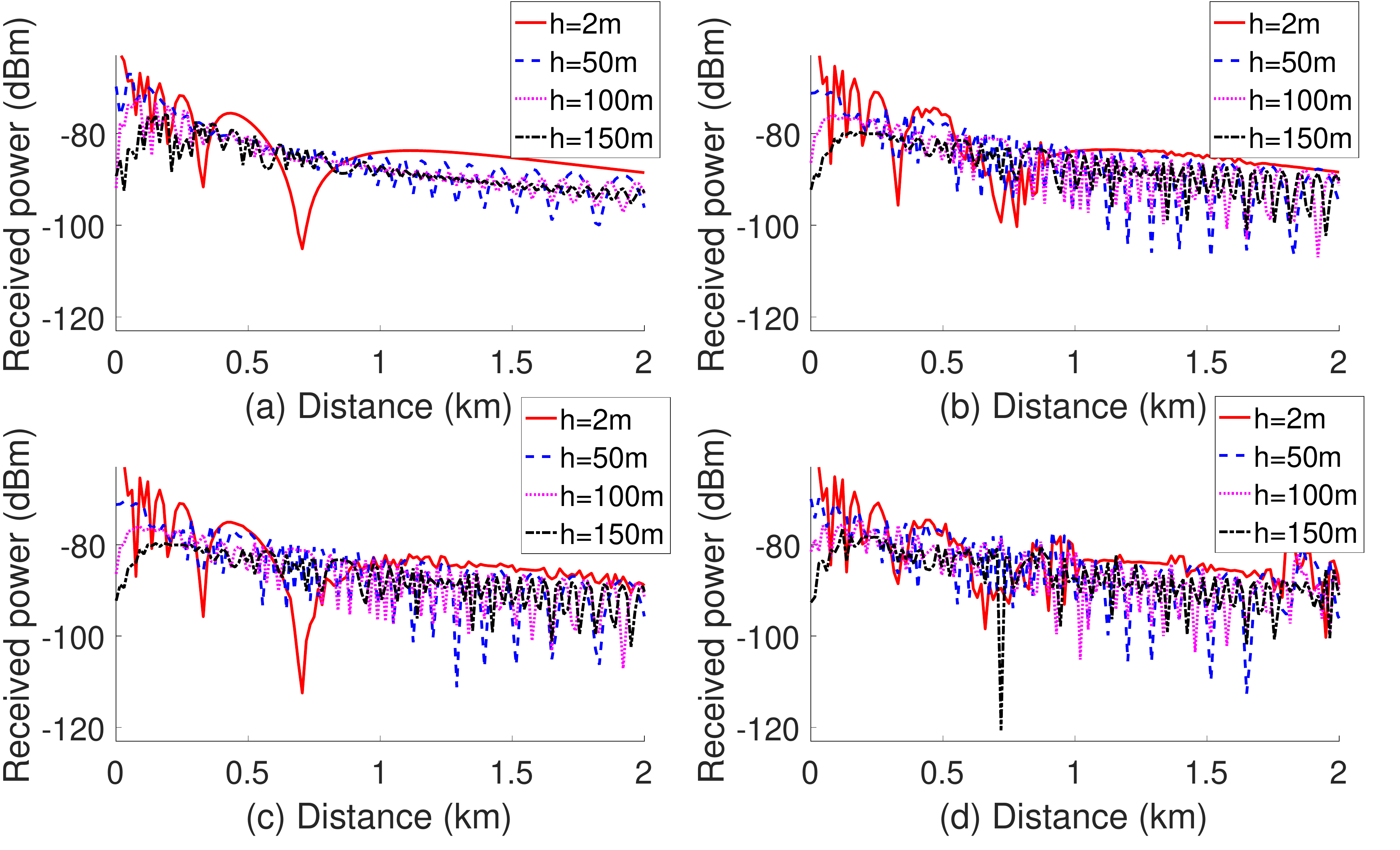}
    \caption{Received power versus distance for different UAV heights at 28 GHz in different scenarios: (a) Over sea, (b) Rural, (c) Suburban, (d) Urban.}\label{Fig:Rec_power_28}
\end{figure}

\begin{figure}[t]
	\centering
	\includegraphics[width=\columnwidth]{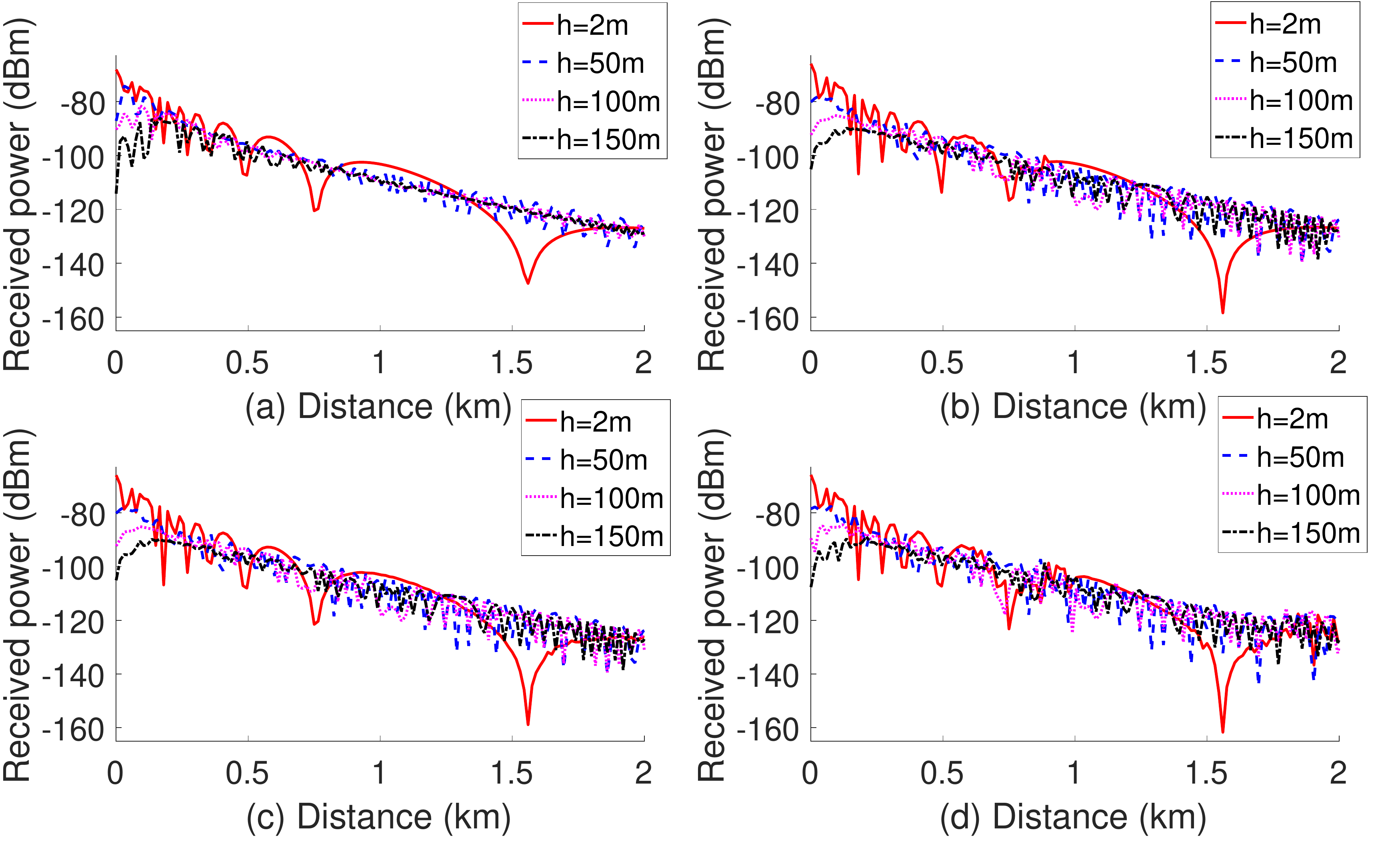}
	    \caption{Received power versus distance for different UAV heights at 60 GHz in different scenarios: (a) Over sea, (b) Rural, (c) Suburban, (d) Urban.}\label{Fig:Rec_power_60}
\end{figure}

\subsection{Ray Tracing Results for the RSS}

The RSS as a function of the distance between the UAV (receiver) and the GS (transmitter) are shown in Fig.~\ref{Fig:Rec_power_28} and Fig.~\ref{Fig:Rec_power_60} for transmission frequencies of 28~GHz and and 60~GHz, respectively. A common observation for RSS in all the scenarios is the two ray propagation model. The presence of scatterers introduces fluctuations in the received power. In case of sub-urban area, the fluctuations result in deviation from the two ray propagation model. At UAV height of 2~m above ground, the two ray propagation model holds only for a specific distance called the critical distance, or first Fresnel zone; after that the received power drops proportionally to some power of the distance~\cite{critical_distance}. The critical distance for 28~GHz and 60~GHz is different due to different RSS decay rates with respect to the distance.        

Results in Fig.~\ref{Fig:Rec_power_28} and Fig.~\ref{Fig:Rec_power_60} show that as the height of the UAV increases, the maxima/minima appearance rate of the two ray propagation model increases as well. Moreover, the maxima/minima appearance rate is higher for 60~GHz compared to 28~GHz. This is due to the smaller wavelength at 60~GHz resulting in constructive and destructive interference more rapidly compared to 28~GHz. Additionally, the rate of RSS decay with respect to distance at 60 GHz is higher than at 28 GHz indicating higher path loss slopes as compared to 28 GHz. The RSS at 60~GHz is lower  compared to 28~GHz as expected, due to higher losses at higher center frequencies.

The selection of relatively low heights in the simulations is due to regulations by the FAA~\cite{FAA} and this limits the effect of the UAV height on the RSS due to scatterers. If the height of the UAV is increased further, the effect of the scatterers on the RSS can be reduced.

In case of over sea scenario, the RSS closely follows the two ray propagation model with minimal fluctuations due to negligible effect of scatterers. On the other hand, in case of rural and suburban scenarios, we observe fluctuations in the RSS near the scatterers. These fluctuations become higher for urban areas due to large number of scatterers as shown in Fig.~\ref{Fig:Rec_power_28} and Fig.~\ref{Fig:Rec_power_60}. An interesting effect is the shadowing of the scatterer reflections, due to the foliage of comparable size to the scatterers weakening the MPCs from scatterers. This can reduce the RSS fluctuations due to weaker reflected MPCs.     



\begin{figure}[t]
	\centering
	\includegraphics[width=\columnwidth]{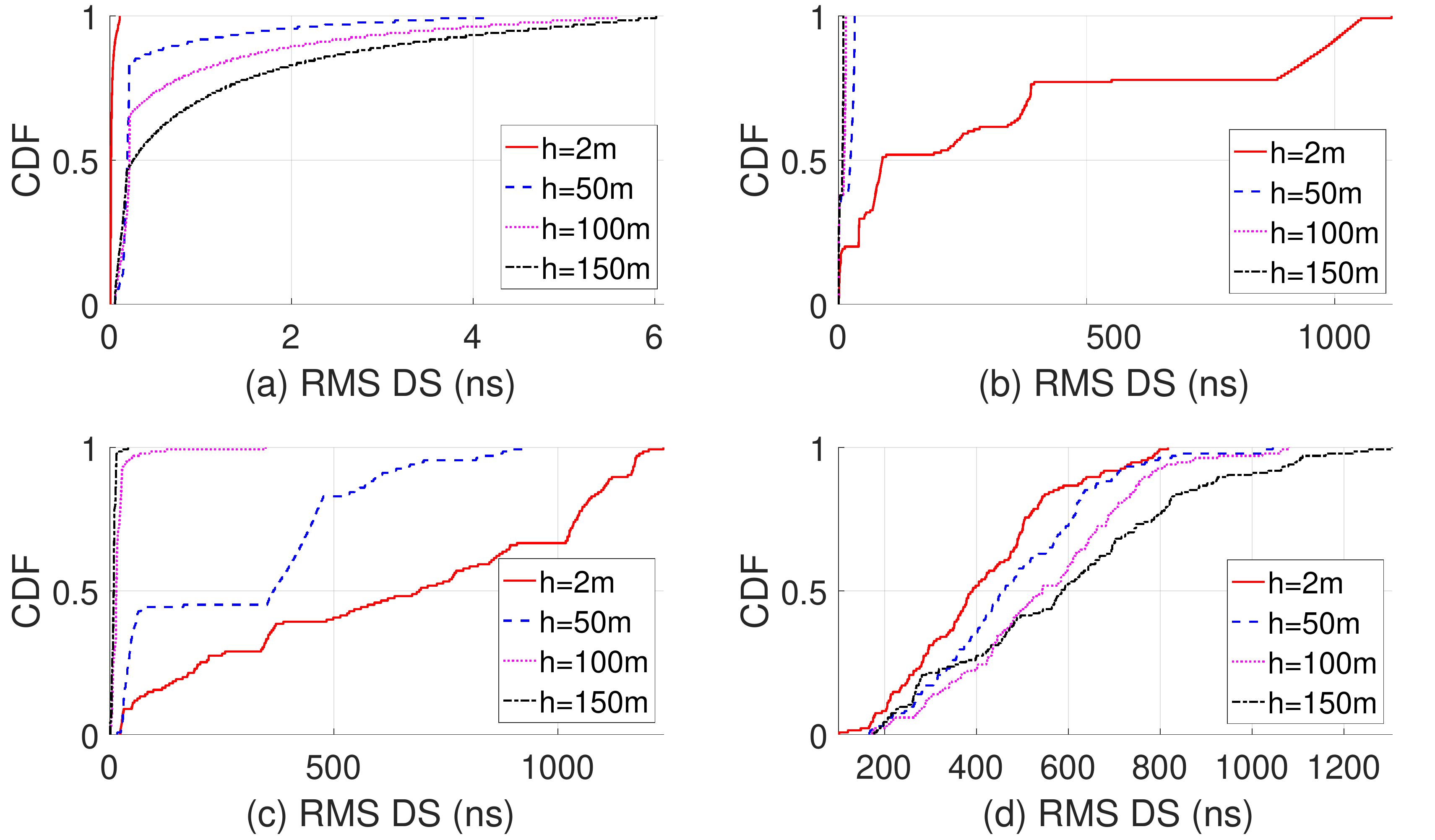}
	    \caption{CDF of RMS-DS for different UAV heights at 28~GHz in different scenarios: (a) Over sea, (b) Rural, (c) Suburban, (d) Urban.}\label{Fig:RMSDS_28}
\end{figure}

\subsection{Ray Tracing Results for the RMS-DS}

\begin{figure}[t]
	\centering
	\includegraphics[width=\columnwidth]{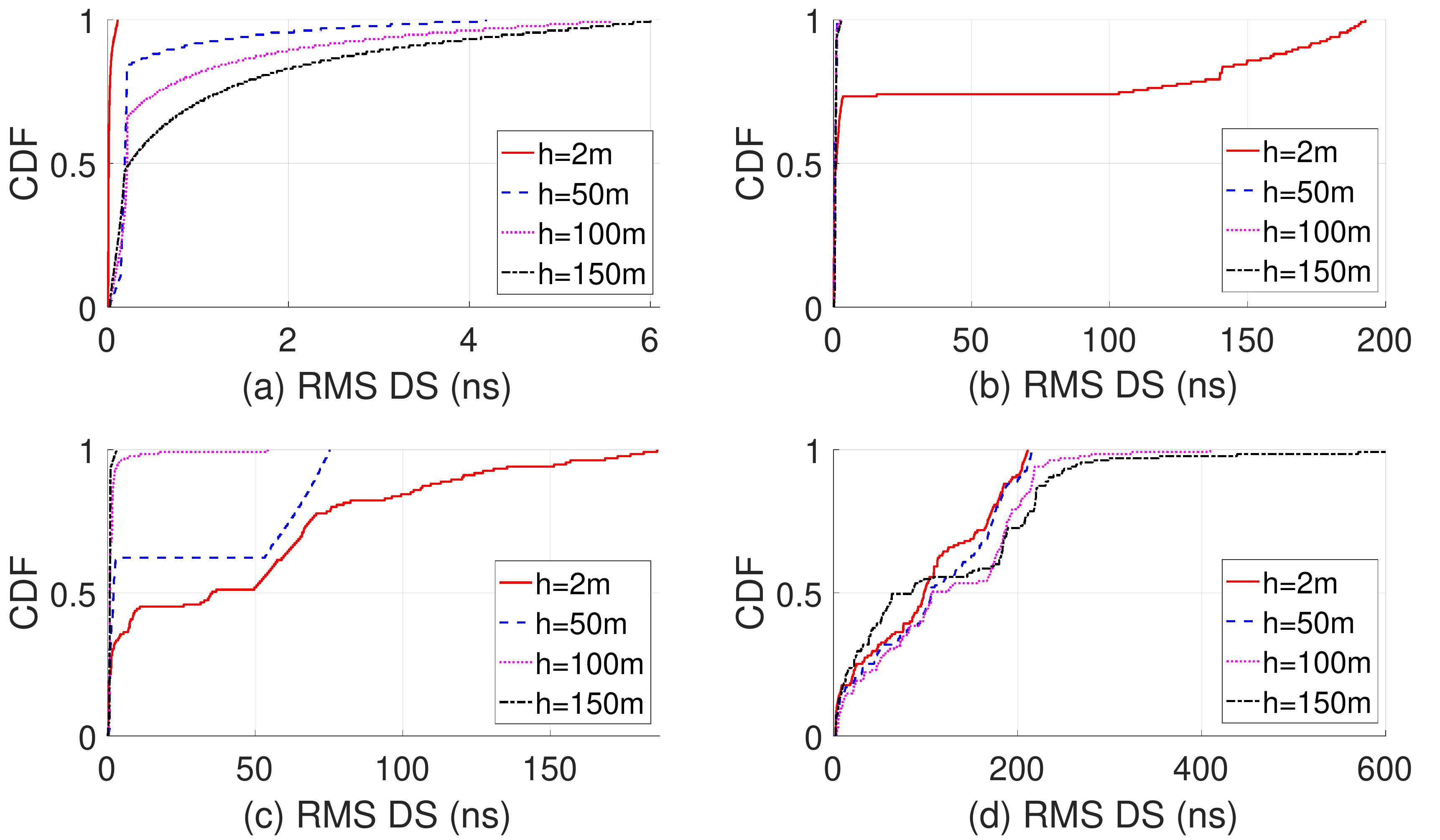}
	    \caption{CDF of RMS-DS for different UAV heights at 60~GHz in different scenarios: (a) Over sea, (b) Rural, (c) Suburban, (d) Urban.}\label{Fig:RMSDS_60}
\end{figure}

In Fig.~\ref{Fig:RMSDS_28} and  Fig.~\ref{Fig:RMSDS_60}, we present the cumulative distribution functions (CDFs) of the RMS-DS of the multipath channel between the GS transmitter and the UAV for four different environments, considering mmWave frequencies of 28~GHz and 60~GHz, respectively. The RMS-DS results in Fig.~\ref{Fig:RMSDS_28} for the mmWave frequency of 28~GHz show that the RMS-DS is the largest for urban environment for most of the UAV heights, due to high density of scatterers in the urban setting. Moreover, we observe that the RMS-DS increases as a function of the UAV height in the urban environment, for the trajectory shown in Fig.~\ref{Fig:setup}(b). The main reason for this behavior is that, at higher UAV altitudes, the UAV moves above the tall buildings, and it can observe signals that are scattered from larger number of surrounding buildings.       

On the other hand, results in Fig.~\ref{Fig:RMSDS_28}(b) and \ref{Fig:RMSDS_28}(c) for rural and suburban environments show that, as opposed to the urban environment, the RMS-DS actually decreases with UAV altitude. In rural/suburban environments, the buildings are not as tall as the urban environment, and more  sparsely deployed. Therefore, at higher UAV altitudes, the signals scattered from the buildings do not arrive at the UAV, hence reducing the RMS-DS. These different behavior of the multipath channel suggest that  environmental factors and UAV height can have significant impact on the channel behavior and hence the receiver design.

The relative impact of different UAV heights on the RMS-DS behavior observed to be similar in Fig.~\ref{Fig:RMSDS_60} (60~GHz mmWave band) when compared with the results in Fig.~\ref{Fig:RMSDS_28} for the 28~GHz band. On the other hand, comparing the results in  Fig.~\ref{Fig:RMSDS_28} and Fig.~\ref{Fig:RMSDS_60} with each other, RMS-DS is lower for the 60~GHz band due to larger path loss at the 60~GHz band.

\section{Channel Sounding for mmWave UAVs}\label{Sec:Sec3}

While ray tracing simulations can provide interesting insights on the behavior of AG mmWave channels as discussed in Section~\ref{Sec:Sec2},  propagation measurements at mmWave bands can more accurately characterize the subtle AG mmWave propagation features. For example, omni-directional RMS-DS with ray tracing simulations in~\cite{zhang2015coverage} for an urban cellular environment have been found be relatively smaller when compared to the RMS-DS for mmWave propagation measurements in similar environments. However, developing a mmWave channel sounder for AG scenarios is a challenging task. Since the available UAVs have limited payload capacity, the basic requirement of a mmWave channel sounder is to be lightweight and compact. In this section, we provide a preliminary framework for AG channel sounding for mmWave UAV channels. 

\subsection{USRP-Based Experimental Setup}\label{Sec:3B}

\begin{figure}[t]
	\centering
	\includegraphics[width=\columnwidth]{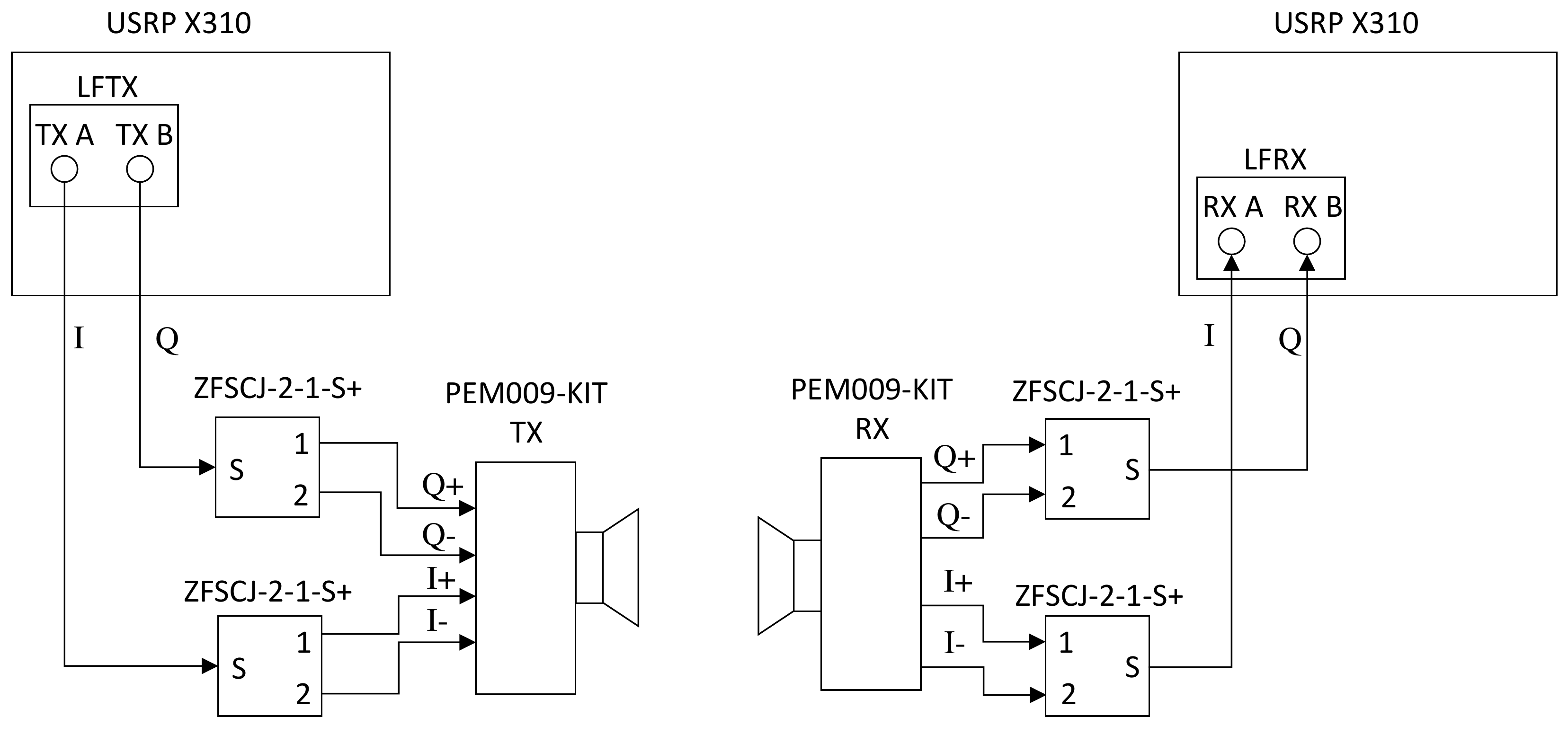}
    \caption{The experimental channel sounder setup.}\label{Fig:setup2}
\vspace{-0.2cm}
\end{figure}

We use 60~GHz Transmit/Receive (TX/RX) development system PEM009-KIT from Pasternack~\cite{pasternack}. This kit accepts differential baseband I/Q signal as an input at the transmitter. The baseband single-ended I/Q signal is generated by using USRP X310 with LFTX daughterboard. This single ended I/Q stream is  translated into differential I/Q stream using two $180^0$ ZFSCJ-2-1-S+ splitters from minicircuits. At the receiving end differential output from the PEM009-KIT is translated into single ended I/Q stream using two ZFSCJ-2-1-S+ splitters as combiners, which is then fed into USRP X310 receiver. The block diagram of the experimental setup is shown in Fig.~\ref{Fig:setup2}.  

\begin{figure}[!t]
	\centering
	\includegraphics[width=0.5\textwidth]{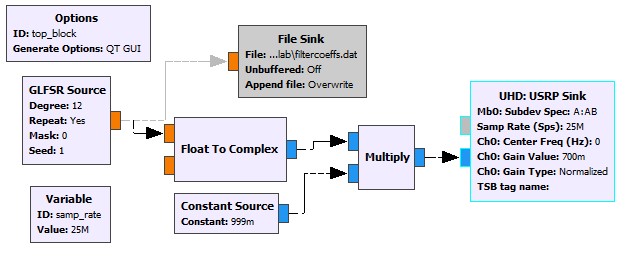}\vspace{-0.3mm}
	\caption{The transmitter GNU radio flow graph for the channel sounder.}\label{Fig:txflow}
\vspace{-0.2 cm}
\end{figure}

In order to sound the channel the GNU-radio flow graph shown in Fig.~\ref{Fig:txflow} is used at the transmitter side. In this flow graph GLFSR source produces a maximal-length Pseudo Noise (PN) sequence $s[n] \in \{ \pm 1 \}$ of degree 12, and of length $2^{12}-1=4095$. We define the sequence $\tilde{s}[n]$ as the periodic extension of $s[n]$ 
obtained by repeating the same sequence indefinitely. This signal is transmitted by the USRP with a sampling rate of 25 MHz after being multiplied by a constant. 


\begin{figure*}[!t]
	\centering
	\includegraphics[width=0.9\textwidth]{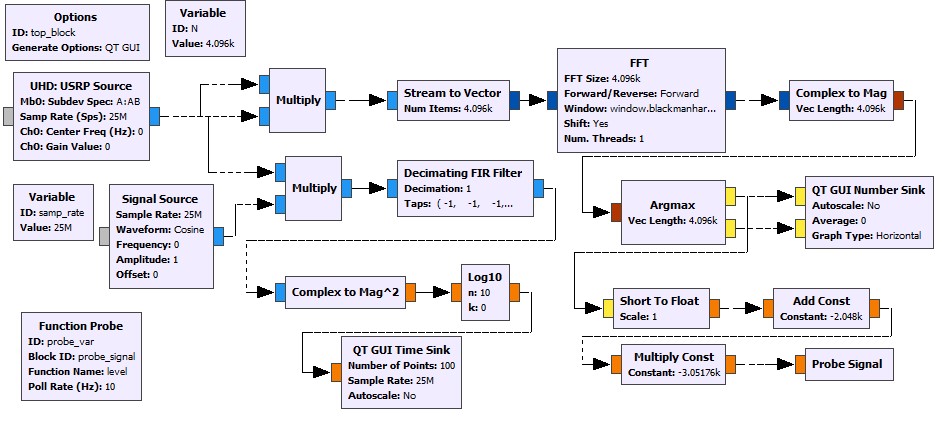}
	\caption{The receiver GNU radio flow graph  for the  channel sounder.}\label{Fig:rxflow}
\end{figure*}

At the receiver side, the USRP source in the flow graph shown in Fig.~\ref{Fig:rxflow} provides samples at $f_{\mathrm{s}}=25$ MHz sampling rate. The carrier frequency offset (CFO) between the transmitter and the receiver is estimated by first taking the square of the received signal. Since the transmitted signal is BPSK modulated, the spectrum of the squared signal contains a dominant peak at two times the value of the  CFO. This estimate for the CFO is then used to compensate for the CFO in the received signal $y(n)$. 
By ignoring the noise, the received signal $y(n)$ can be expressed as the result of the signal $\tilde{s}(n)$ passing through the wireless multipath propagation channel of $N_{\rm P}$ MPCs as follows~\cite{albornoz2001wideband}
\begin{equation}
y(n)=\sum_{p=0}^{N_{\rm P}-1} a_p \exp(j\theta_p) \tilde{s}(n-\tau_p).
\end{equation}
Since $s[n]$ can be assumed known at the receiver side, a matched filter, $g(n)=s(-n)$, can be applied after CFO correction to obtain
\begin{eqnarray}
r(n)&=& y(n) * g(n) \\
&=& \sum_{p=0}^{N_{\rm P}-1} a_p \exp(j\theta_p) \tilde{R}_{{s}}(n-\tau_p),
\end{eqnarray}
where $\tilde{R}_{{s}}(n)$ is periodic extension of deterministic auto-correlation function of $s(n)$. The signal $r(n)$ is composed of $N_{\rm P}$ attenuated, phase shifted, and delayed copies of the autocorrelation function, and it represents the channel impulse response (CIR) for a given environment. 

\section{Conclusions}

In this paper, we provide ray tracing simulations for mmWave UAV AG propagation channels. It is observed that the two ray propagation model can be applicable (with some limitations) for urban, suburban, rural, and over sea scenarios. The presence of scatterers affect the RSS especially in case of urban scenario. The fluctuation rate of the RSS with respect to distance at~60GHz is higher when compared with 28~GHz. The RMS-DS behavior is highly dependent on the height of the UAV as well as the density/height of the scatters around the UAV. 
For the suburban and rural scenarios, we observe a lower RMS-DS when the UAV height increases, due to less significant effect of the scatterers at higher UAV heights. We also provide an experimental setup for AG mmWave channel sounding measurements using USRPs and GNU radio. The USRP-based channel sounder is lightweight enough to be carried, for example, at a DJI S-1000 octocopter, for conducting channel sounding experiments at 60~GHz spectrum. 











\bibliographystyle{IEEEtran}





\end{document}